\documentclass[11pt]{article}
\usepackage{hyperref}
\pdfoutput=1

\begin{document}

\title{Turbulent flow over a house in a \\ simulated hurricane boundary layer}

\author{Zachary Taylor$^{1}$, Murray Morrison$^{1}$, Roi Gurka$^{2}$, Gregory Kopp$^{1}$ \\
\\\vspace{6pt} $^{1}$Civil and Environmental Engineering, \\ University of Western Ontario, Canada \\
\\\vspace{6pt} $^{2}$Chemical Engineering, \\ Ben-Gurion University, Israel}

\maketitle

\begin{abstract}
Every year hurricanes and other extreme wind storms cause billions of dollars in damage worldwide. For residential construction, such failures are usually associated with roofs, which see the largest aerodynamic loading. However, determining aerodynamic loads on different portions of North American houses is complicated by the lack of clear load paths and non-linear load sharing in wood frame roofs. This problem of fluid-structure interaction requires both wind tunnel testing and full-scale structural testing.\\
A series of wind tunnel tests have been performed on a house in a simulated atmospheric boundary layer (ABL), with the resulting wind-induced pressures applied to the full-scale structure.  The ABL was simulated for flow over open country terrain where both velocity and turbulence intensity profiles, as well as spectra, were matched with available full scale measurements for this type of terrain.  The first set of measurements was 600 simultaneous surface pressure measurements over the entire house.\\
A key feature of the surface pressure field is the occurrence of large, highly non-Gaussian, peak uplift (suctions) on the roof.  In order to better understand which flow features cause this, PIV experiments were performed on the wind tunnel model.  These experiments were performed with time-resolved PIV (sampling rate of 500 Hz) for a duration of 30 seconds.  From the fluid dynamics videos (\href{http://ecommons.library.cornell.edu/bitstream/1813/14089/3/taylor_gofm2009_mpeg1.mpg}{low-} and \href{http://ecommons.library.cornell.edu/bitstream/1813/14089/2/taylor_gofm2009_mpeg2.mpg}{high-}resolution) generated from the PIV data it is clear that strong circulation is generated at the windward edge of the roof.  These vortices are eventually shed and convect along the roof.  It is the presence of this concentrated circulation which is responsible for the peak loading observed.
\end{abstract}

\end{document}